\documentclass[aps,pop,showpacs,twocolumn,superscriptaddress,letterpaper]{revtex4-1}
\usepackage{mathrsfs}
\usepackage{amsmath}
\usepackage{bm}
\usepackage{color}
\usepackage{graphicx}
\usepackage{hyperref}
\usepackage{float}
\usepackage{listings}
\hypersetup{pdfpagemode=UseNone,pdfstartview=FitH,pdfstartpage=1}

\begin{document}

\title{Rapid Computation of the Plasma Dispersion Function: Rational and Multi-pole Approximation, and Improved Accuracy}

\author{Huasheng Xie}
\email[]{Email: huashengxie@gmail.com, xiehuasheng@enn.cn} \affiliation{Hebei Key Laboratory of Compact Fusion, Langfang 065001, China}
\affiliation{ENN Science and Technology Development Co., Ltd., Langfang 065001, China}

\date{\today}

\begin{abstract}
The plasma dispersion function $Z(s)$ is a fundamental complex special integral function widely used in the field of plasma physics. The simplest and most rapid, yet accurate, approach to calculating it is through rational or equivalent multi-pole expansions. In this work, we summarize the numerical coefficients that are practically useful to the community. Besides the Padé approximation to obtain coefficients, which are accurate for both small and large arguments, we also employ optimization methods to enhance the accuracy of the approximation for the intermediate range. The best coefficients provided here for calculating $Z(s)$ can deliver twelve significant decimal digits. This work serves as a foundational database for the community for further applications.
\end{abstract}


\maketitle

\section{Introduction and motivation}\label{sec:intro}
The plasma dispersion function\cite{Fried1961,Huba2009,Gurnett2005} is defined as
\begin{equation}\label{eq:Z}
Z(s)=\frac{1}{\sqrt{\pi}}\int_{-\infty}^{\infty}\frac{e^{-t^2}}{t-s}dt,
\end{equation}
with $s=x+iy$, which is valid for $y>0$. For $y\leq0$, the function is analytically continued from the above upper plane form to the lower plane. The function is relevant to the Faddeev function $w(s)$ and error function ${\rm erf}(s)$ by
\begin{eqnarray}
Z(s)&=&i\sqrt{\pi} w(s),\\
Z(s)&=&i\sqrt{\pi}e^{-s^2}[1+{\rm erf}(is)]\\
&=&i\sqrt{\pi}e^{-s^2}[1+i\cdot{\rm erfi}(s)]\\
&=&i\sqrt{\pi}e^{-s^2}-2F(s),
\end{eqnarray}
where $i=\sqrt{-1}$, ${\rm erf}(s)=\frac{2}{\sqrt{\pi}}\int_0^se^{-t^2}dt$ and ${\rm erfi}(s)=-i\cdot{\rm erf}(is)$. Here, $F(x)$ is Dawson integral $F(x)=e^{-x^2}\int_0^xe^{t^2}dt=\frac{\sqrt{\pi}}{2}e^{-x^2}\cdot{\rm erfi}(x)$. Note also $dZ/ds=-2[1+sZ(s)]$.
The below symmetry properties (the asterisk denotes complex conjugation) hold
\begin{eqnarray}
Z(s)=-[Z(-s^*)]^*,\\\label{eq:sys}
Z(s)=[Z(s^*)]^*+2i\sqrt{\pi}\exp[-s^2]~~(y<0).
\end{eqnarray}
And the two-side Taylor expansion approximation
\begin{eqnarray}\label{eq:Ztaylor}
&&Z(s)\simeq\\\nonumber
&&\left\{
\begin{aligned}
    \sum_{k=0}^{\infty}a_ks^k \simeq i\sqrt{\pi}e^{-s^2}-s\sum_{n=0}^{\infty}(-s^2)^n\frac{\Gamma(1/2)}{\Gamma(n+3/2)},~s\to0 \\
    \sum_{k=0}^{\infty}a_{-k}s^{-k} \simeq i\sigma\sqrt{\pi}e^{-s^2}-\sum_{n=0}^{\infty}\frac{\Gamma(n+1/2)}{\Gamma(1/2)s^{2n+1}},~s\to\infty
\end{aligned}
\right.
\end{eqnarray}
where
\begin{eqnarray}
\sigma=\left\{
\begin{aligned}
   & 0,~~y>0, \\
   &1,~~y=0, \\
   & 2,~~y<0,
\end{aligned}
\right.
\end{eqnarray}
and $\Gamma$ is Euler's Gamma function. A further expansion is $e^{-s^2}=\sum_{n=0}^{\infty}\frac{(-s^2)^n}{n!}$. If term $i\sigma\sqrt{\pi}e^{-s^2}$ is omitted, Eq.(\ref{eq:Ztaylor}) would not match well of the imag part for the range $y<\sqrt{\pi}x^2e^{-x^2}$ when $x\gg1$. Since $e^{-x^2}<10^{-40}$ for $x>10$, the term actually mainly affects the intermediate range, e.g., $1<x<10$.

The direct numerical integral of Eq.(\ref{eq:Z}) would be time-consuming and inaccurate due to the pole in the integral dominator. Series methods, cf., \cite{Martin1980,Hui1978,Humlicek1979,Humlicek1982,Weideman1994,Zaghloul2011,Franklin1968,Nemeth1981,
Fried1968,Newberger1986,Tjulin2000,Ronnmark1982,Xie2016,Xie2019}, have been proposed to numerically calculate $Z(s)$. A comparison of different methods is also provided \cite{Zaghloul2011}. The yet simplest and fastest, yet still accurate, approach to calculate it is to use rational expansion \cite{Hui1978,Humlicek1979,Humlicek1982,Hunana2019} or equivalent multi-pole expansion \cite{Fried1968,Martin1980,Ronnmark1982,Xie2016,Xie2019}.

It is also surprising that the rational and multi-pole approximation of the plasma dispersion function inspire two unexpected applications: (1) The development of Landau fluid models to mimic the kinetic Landau damping effects \cite{Hammett1990,Hammett1992,Hunana2019}; (2) The first solver to obtain all the kinetic dispersion relation solutions without the requirement of an initial guess \cite{Xie2016,Xie2019}. These two applications are possible only when the approximation of $Z(s)$ keeps the following features: (a) The approximation should be rational functions; (b) One formula for all the interesting regions; (c) To maintain the same accuracy with as few terms as possible.

Hence, segmentation calculations (cf., \cite{Zaghloul2011}) and non-rational expansion are not our choices. We find our only choice is the rational and multi-pole approximation. The standard approach to obtaining the rational coefficients is to use Pade approximation to match Eq.(\ref{eq:Ztaylor}), which is accurate for small ($s\to0$) and large ($s\to\infty$) arguments. The coefficients can be calculated rigorously via matrix inverse. The Pade approximation is less accurate at the intermediate range $s\simeq1$. We then use optimization methods to reduce the error of the approximation at the intermediate range, which can improve one order of magnitude. Weideman's \cite{Weideman1994} method is also in rational form; however, it will require series of high-order terms.

We found rational and multi-pole coefficients are not provided systematically in literature. The methods used in this work are probably not new, and some of the coefficients have been provided in different literature, for example, the Pade rational form with also analytic coefficients up to $J=2-8$ \cite{Martin1980,Hunana2019}, multi-pole for small $J\leq8$ \cite{Huba2009,Martin1980,Ronnmark1982,Xie2016} and extended to $J=24$ \cite{Xie2019}. The coefficients with optimization for $J=5,6,7$ can also be found in rapid calculation of $Z$, such as Refs. \cite{Hui1978,Humlicek1979,Humlicek1982}. A systematic summary of the rational coefficients and corresponding multi-pole coefficients could be useful to the community, especially for beginners.

The purpose of the present work is to provide comprehensive numerical Pade coefficients from small order to high accuracy, typically $J=2$ to $J=24$, which can have an error less than $10^{-13}$, and could be used for rapid numerical calculation of $Z(s)$ to high accuracy. The coefficients of improved fitting for small $J\leq8$ are also provided, which can be used to save computation time or improve accuracy with fewer terms in Landau fluid model and kinetic dispersion relation solver.

In Section \ref{sec:approaches}, we describe the approaches we used and the results we obtained. In Section \ref{sec:summ}, summary and conclusion are given.

\section{Approaches and Results}\label{sec:approaches}

The rational approximation and multi-pole expansion of $Z(s)$ is (typos in Ref. \cite{Xie2016} are fixed here)
\begin{eqnarray}\label{eq:ZAform}
Z(s)\simeq Z_A^J(s)=\frac{\sum_{l=0}^{J-1}p_ls^l}{q_0+\sum_{k=1}^Jq_ks^k}=\sum_{j=1}^J\frac{b_j}{s-c_j},
\end{eqnarray}
with $q_0=1$. This form is valid for the upper plane to high accuracy and is analytical (i.e., automatically be analytically continued) and thus would also have a good approximation at the real axis and lower plane in case the $y$ is not far from the real axis. If one needs a more accurate value in the lower plane, Eq.(\ref{eq:sys}) can be used to use the value from the upper plane and symmetry property. Hence, the entire plane can be calculated to high accuracy.

\subsection{Pade method}
The Pade expansion is to match Eq.(\ref{eq:ZAform}) to Eq.(\ref{eq:Ztaylor}), i.e., with terms to terms match
\begin{eqnarray}\label{eq:Padematch}
\Big[q_0+\sum_{k=1}^Jq_ks^k\Big]Z(s)=\sum_{l=0}^{J-1}p_ls^l.
\end{eqnarray} 
To obtain the coefficients, the system of equations to be solved are
\begin{subequations}\label{eq:Padeeq}
\begin{align}
p_j=\sum_{k=0}^{j}a_kq_{j-k},~~1\leq j\leq I,\\
p_{J-j}=\sum_{k=0}^{j}a_{-k}q_{J+k-j},~~1\leq j\leq K,
\end{align}
\end{subequations}
where $I+K=2J$, and $p_j=0$ for $j>J-1$ and $j<0$, and $q_j=0$ for $j>J$ and $j<0$. The $2J$ equations determine $2J$ coefficients $p_j$ and $q_j$. Here $I$ means keeping $I$ equations for $s\to0$, and $K$ means keeping $K$ equations for $s\to\infty$. The above equation can be solved via matrix inverse, both analytically for small $J$ or numerically for arbitrary $J$. 

The term $i\sigma\sqrt{\pi}e^{-s^2}$ for $s\to\infty$ in Eq.(\ref{eq:Ztaylor}) is omitted, and can be explicitly rewritten as
\begin{eqnarray}\label{eq:Ztaylor1}
&&Z(s)\simeq\\\nonumber
&&\left\{
\begin{aligned}
    \sum_{k=0}^{\infty}a_ks^k \simeq& i\sqrt{\pi}-2s-i\sqrt{\pi}s^2+\frac{4}{3}s^3+\\
&\frac{i\sqrt{\pi}}{2}s^4-\frac{8}{15}s^5+\cdots,~s\to0 \\
    \sum_{k=0}^{\infty}a_{-k}s^{-k} \simeq& -\frac{1}{s}-\frac{1}{2s^3}-\frac{3}{4s^5}-\frac{15}{8s^7}+\cdots,~s\to\infty
\end{aligned}
\right.
\end{eqnarray}
And Eq.(\ref{eq:Padeeq}) is rewritten as
\begin{subequations}\label{eq:Padeeq1}
\begin{align}
p_0=i\sqrt{\pi},\\
p_1=-2+i\sqrt{\pi}q_1,\\
p_2=-i\sqrt{\pi}-2q_1+i\sqrt{\pi}q_2,\\
\cdots,\\
-q_{J}=p_{J-1},\\
-q_{J-1}=p_{J-2},\\
-q_{J-2}-\frac{1}{2}q_J=p_{J-3},\\
\cdots.
\end{align}
\end{subequations}

For a given $J$ and $I$, the coefficients $p_l$ and $q_k$ can be solved easily via matrix inverse. Solving for the multi-pole coefficients $b_j$ and $c_j$ is also straightforward (e.g., using the residue() function in Matlab), i.e., $c_j$ are the roots of the equation $q_0+\sum_{k=1}^Jq_ks^k=0$.
There are also symmetric features to ensure that $b_j$ and $c_j$ occur in pairs: $b_j=b_{J+1-j}^*$ and $c_j=-c_{J+1-j}^*$.

For multi-pole expansion, we have
\begin{eqnarray}\label{eq:ZtaylorJpole}
Z_A(s)\simeq\sum_{j=1}^Jb_j\left\{
\begin{aligned}
    -\frac{1}{c_j}-\frac{s}{c_j^2}-\frac{s^2}{c_j^3}+\cdots,~s\to0 \\
   \frac{1}{s}+\frac{c_j}{s^2}+\frac{c_j^2}{s^3}+\cdots,~s\to\infty
\end{aligned}
\right.
\end{eqnarray}
Comparing Eq.(\ref{eq:ZtaylorJpole}) with Eq.(\ref{eq:Ztaylor1}), we have $\sum_jb_j/c_j=-i\sqrt{\pi}$, $\sum_jb_j/c_j^2=2$, $\sum_jb_j/c_j^3=i\sqrt{\pi}$, and $\sum_jb_j=-1$, $\sum_jb_jc_j=0$, and $\sum_jb_jc_j^2=-1/2$. For kinetic dispersion relation solver \cite{Xie2016,Xie2019}, $\sum_jb_jc_j^2=-1/2$ is used, which means that we need to keep to $O(1/s^3)$ when calculating the multi-pole coefficient. Hence, we should have $K\geq3$. This also implies that $-q_{J}=p_{J-1}$, $-q_{J-1}=p_{J-2}$, and $-q_{J-2}-\frac{1}{2}q_J=p_{J-3}$ hold, and the $R(s)=1+sZ(s)$ coefficients in the Landau fluid model can be obtained straightforwardly. For small $J=2,3,4$, we may only keep to $O(1/s)$ or $O(1/s^2)$. The coefficients taken by Ronnmark \cite{Ronnmark1982} are $J=8$, $I=10$.

It also appears that a slightly larger $I$ \cite{Martin1980} than $K$ can provide a better overall approximation. The notation $Z_{IK}$ is used to describe different orders of approximation.

\begin{figure*}
\centering
\includegraphics[width=15cm]{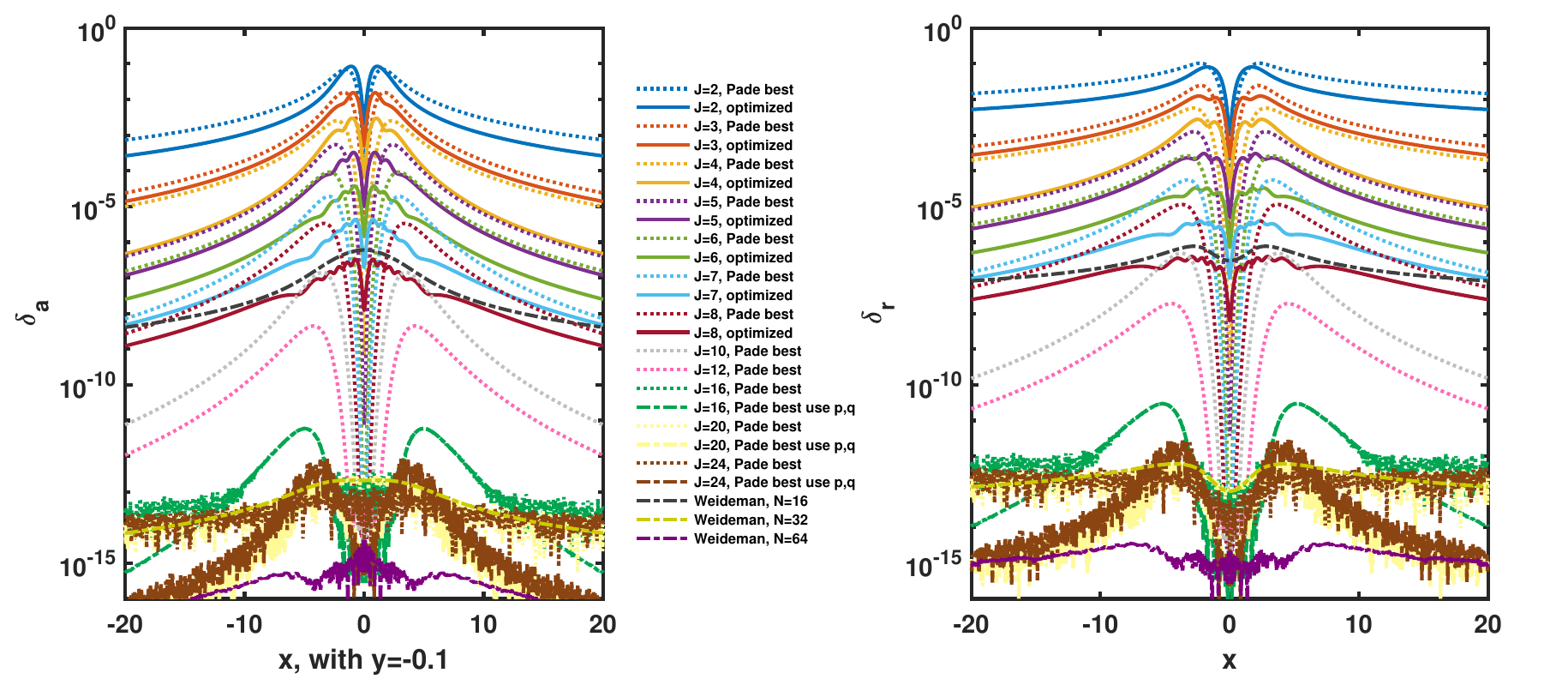}
\caption{Errors of $Z$ using different $J$ poles coefficients and Weideman coefficients, with $y=-0.1$.}\label{fig:zJpoleerr_y=-0.1}
\end{figure*}

\begin{figure*}
\centering
\includegraphics[width=15cm]{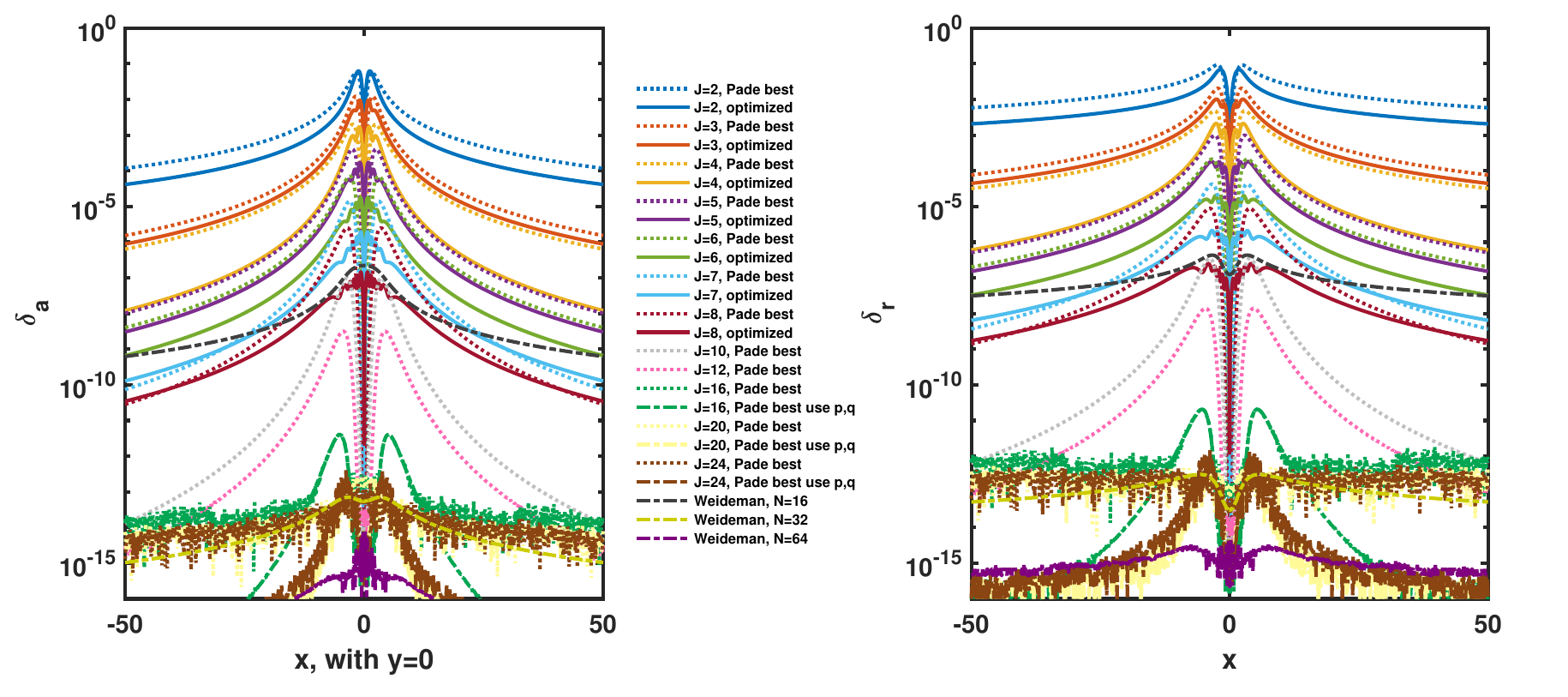}
\caption{Errors of $Z$ using different $J$ poles coefficients and Weideman coefficients, with $y=0$.}\label{fig:zJpoleerr_y=0}
\end{figure*}

\begin{figure*}
\centering
\includegraphics[width=15cm]{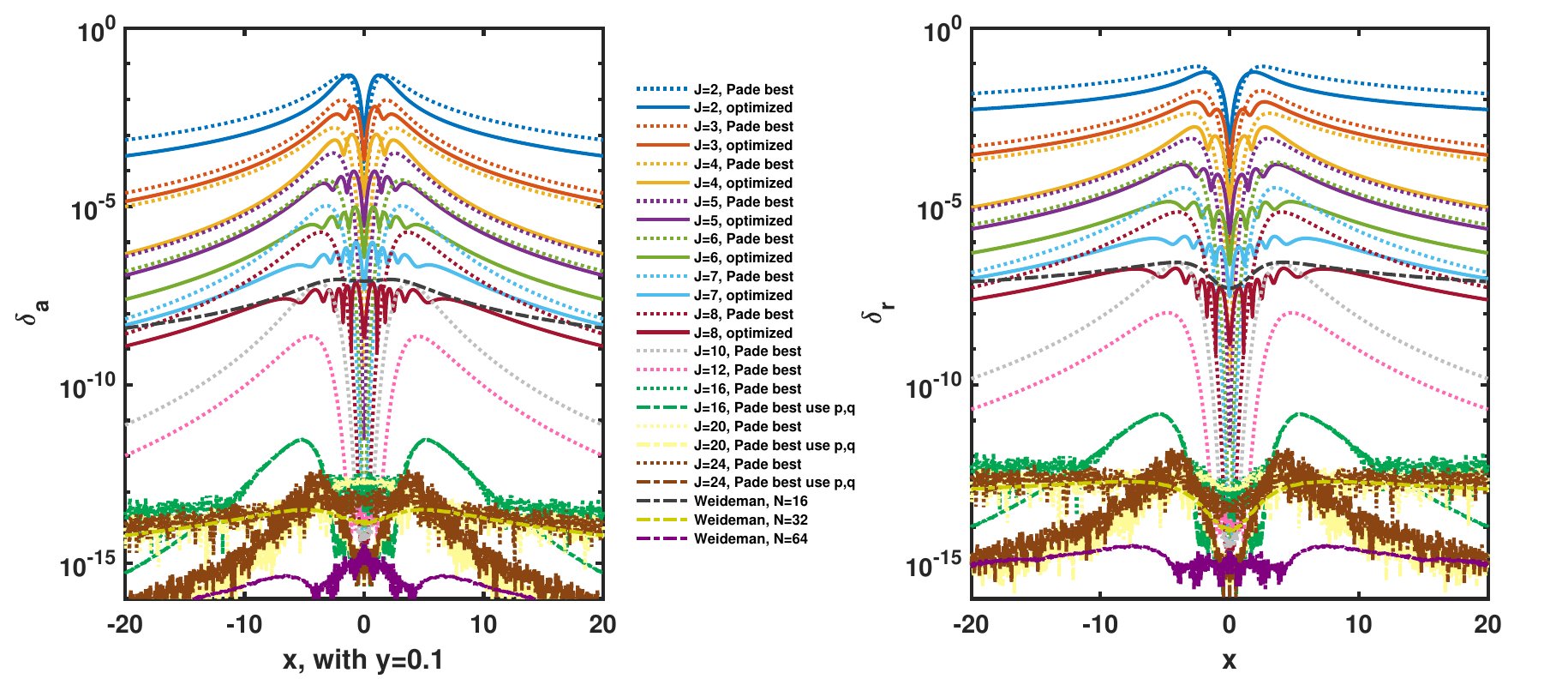}
\caption{Errors of $Z$ using different $J$ poles coefficients and Weideman coefficients, with $y=0.1$.}\label{fig:zJpoleerr_y=0.1}
\end{figure*}

\subsection{Search minimum method}
It appears that $p_{2j}$ and $q_{2j+1}$ are pure imaginary numbers, and $p_{2j+1}$ and $q_{2j}$ are pure real numbers, for $j=0,1,2,\cdots$. 

We minimize both the absolute and relative errors by
\begin{eqnarray}\label{eq:minZ}
min\Big\{w\delta_a^2+(1-w)\delta_r^2\Big\},
\end{eqnarray}
with
\begin{eqnarray}\label{eq:minZ}
\delta_a=|Z_A(s)-Z(s)|,~~~\delta_r=|Z_A(s)/Z(s)-1|,
\end{eqnarray}
and $Z_A(s)$ is the approximation value, and $Z(s)$ is the accurate value. In this work, we set $w=0.5$. We take the accurate $Z(s)$ to be the one using the Dawson function in Matlab (though it may not be accurate for all ranges). We use some of the constraint Eq.(\ref{eq:Padeeq1}) to make small $s$ to $O(s^2)$ and large $s$ to $O(1/s^3)$. There are some standard approaches to obtain the optimized $p_l$ and $q_k$. Here, we use the Matlab function \texttt{fminsearch()} to perform the calculations and use the Pade coefficients as initial guesses. The results could be sensitive to the initial guess and the iterative convergence criteria. Hence, the final results may not be determined. We choose our best obtained results to list here. The optimization is performed for $s=x+iy$, with $x=[-50,50]$ and $y=0.1$.

\begin{figure*}
\centering
\includegraphics[width=15cm]{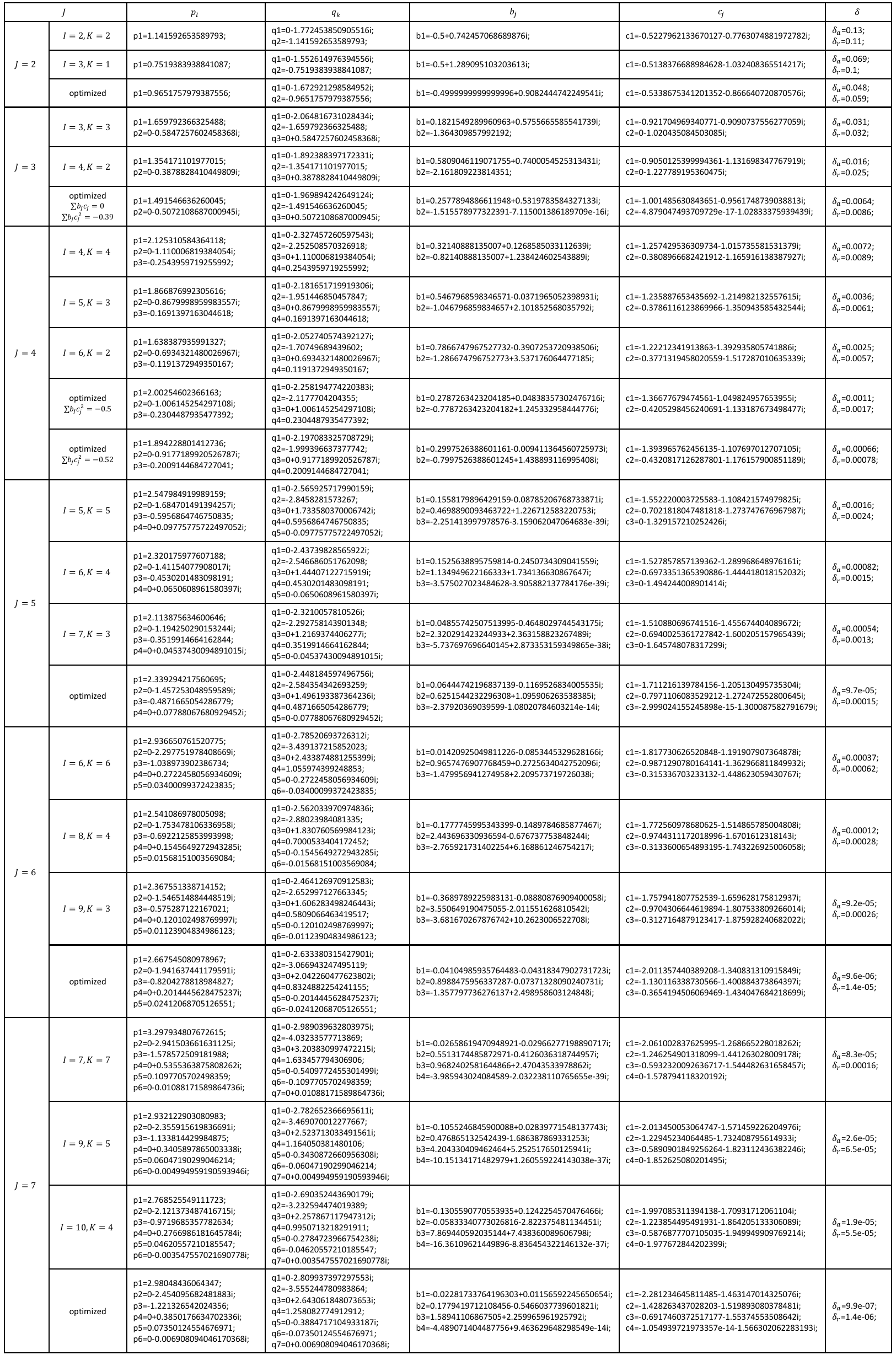}
\caption{Coefficients $p_l$, $q_k$, $b_j$, $c_j$ and corresponding errors $\delta_a$ and $\delta_r$ for $J=2-7$. For Pade $K\geq3$ and for optimized $J\geq4$, we ensure $|\sum_jb_jc_j^2+1/2|<10^{-12}$. Note also $p_0=i\sqrt{\pi}$, $b_j=b_{J+1-j}^*$ and $c_j=-c_{J+1-j}^*$.}\label{fig:coef_J2to7}
\end{figure*}

\begin{figure*}
\centering
\includegraphics[width=15cm]{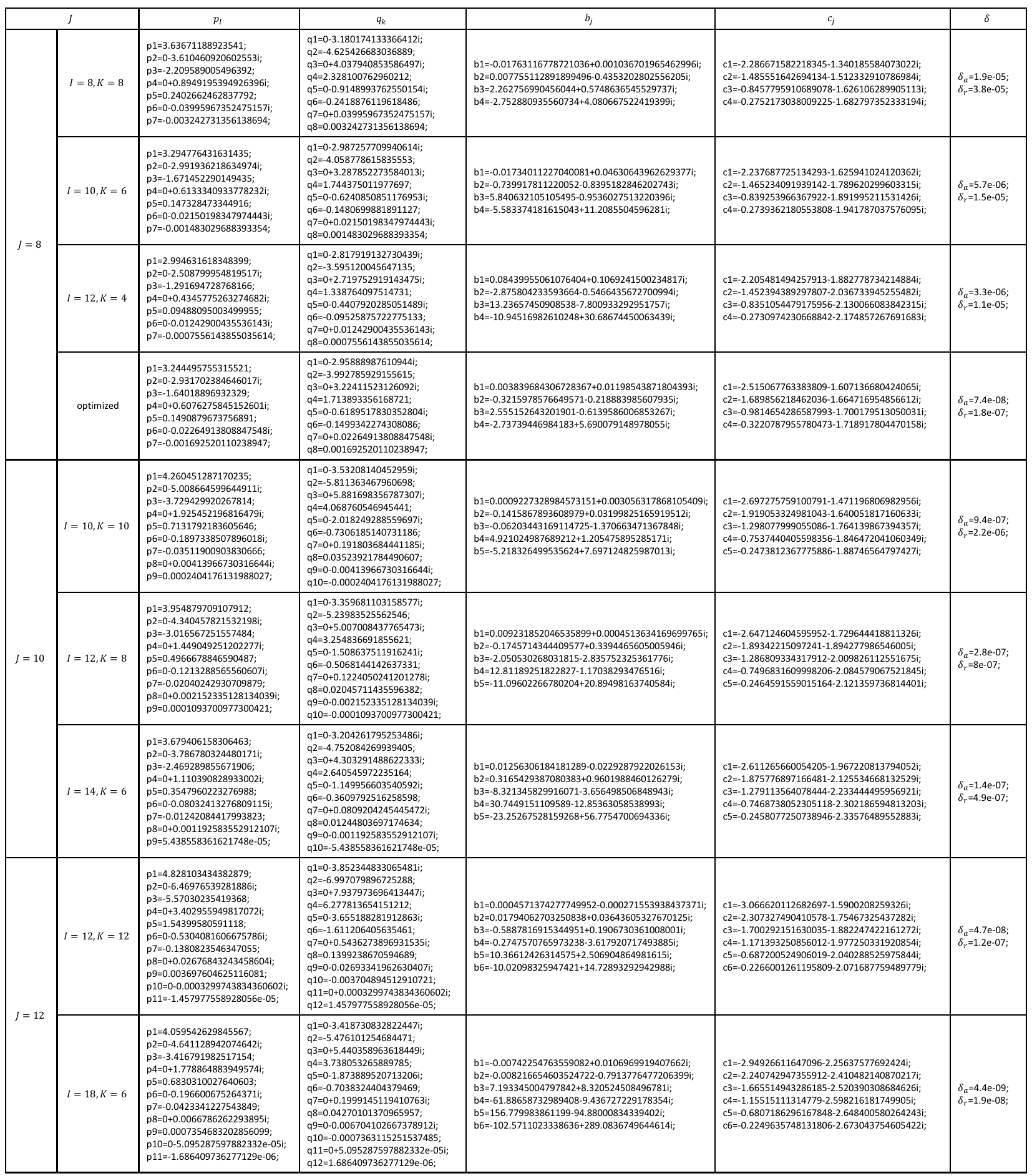}
\caption{Coefficients $p_l$, $q_k$, $b_j$, $c_j$ and corresponding errors $\delta_a$ and $\delta_r$ for $J=8,10,12$.  Note also $p_0=i\sqrt{\pi}$, $b_j=b_{J+1-j}^*$ and $c_j=-c_{J+1-j}^*$.}\label{fig:coef_J8to12}
\end{figure*}

\begin{figure*}
\centering
\includegraphics[width=15cm]{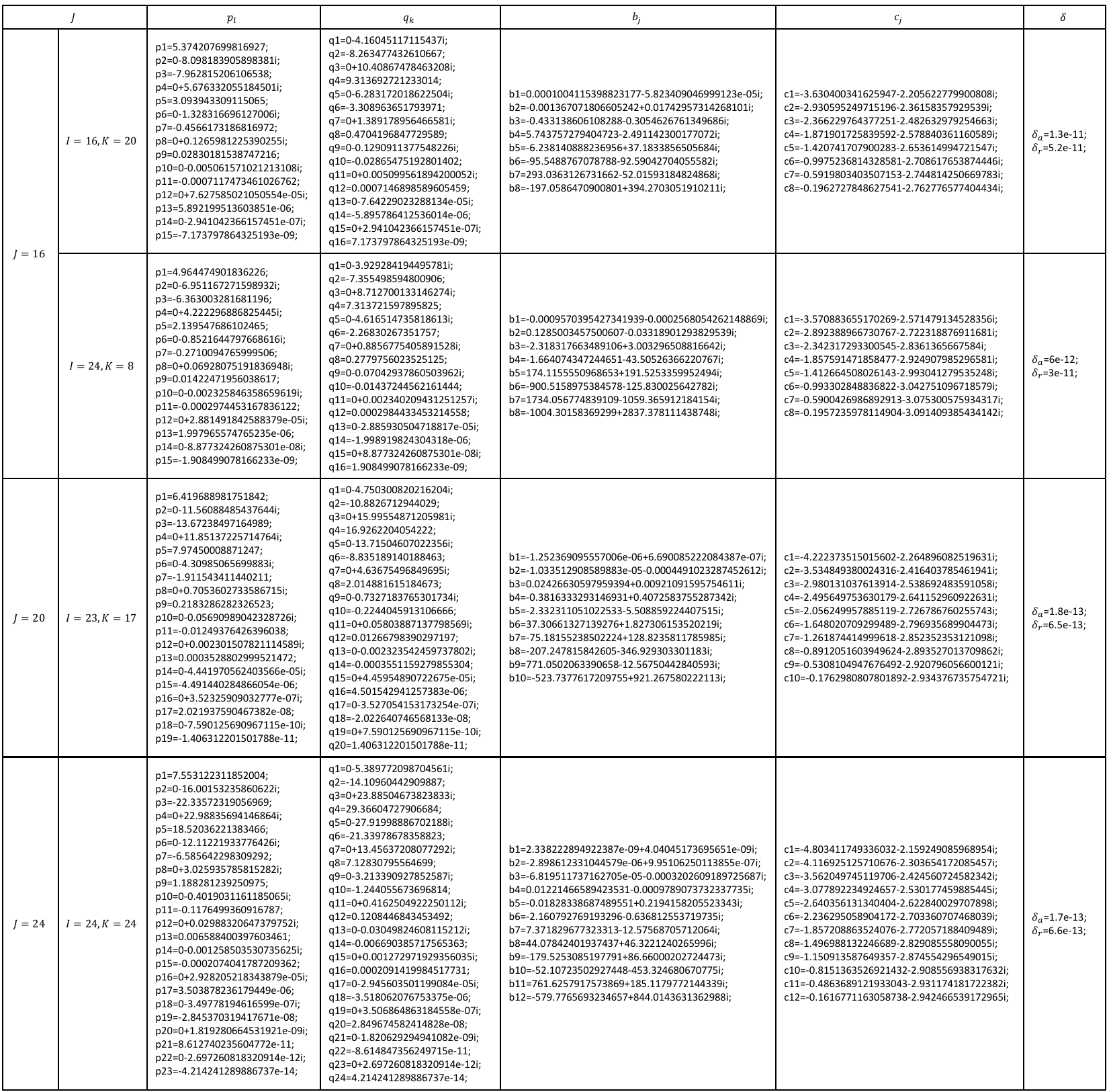}
\caption{Coefficients $p_l$, $q_k$, $b_j$, $c_j$ and corresponding errors $\delta_a$ and $\delta_r$ for $J=16,20,24$.  Note also $p_0=i\sqrt{\pi}$, $b_j=b_{J+1-j}^*$ and $c_j=-c_{J+1-j}^*$.}\label{fig:coef_J16to24}
\end{figure*}

\subsection{Coefficients tables}
After the methods described in the subsections above, we performed calculations to obtain the coefficients $p_l$, $q_k$, $b_j$, $c_j$, and corresponding errors $\delta_a$ and $\delta_r$ in a table. For Pade coefficients, we calculated from $J=2$ to $24$, and from $J=2$ to $8$ for optimized coefficients. The error data are taken for the lower plane with $y=-0.1$ and $x=[-50,50]$, ensuring that the approximation can also accurately capture the Landau damping effect even without the $2i\sqrt{\pi}e^{-s^2}$ term. Data for $J>24$ are not listed here, as the double precision is not adequate and not necessary for most applications.

Figures \ref{fig:coef_J2to7}, \ref{fig:coef_J8to12}, and \ref{fig:coef_J16to24} show the coefficients, while Figures \ref{fig:zJpoleerr_y=-0.1}, \ref{fig:zJpoleerr_y=0}, and \ref{fig:zJpoleerr_y=0.1} show the results of the errors. `Pade best' in the figures means that $I$ is chosen such that the error is minimized according to the tables.

It is possible to optimize for $J\geq9$, but it is less useful, and thus we do not provide it here. The reason is that $J=8$ is sufficient for most purposes, and those requiring higher accuracy can use slightly higher $J$ Pade coefficients; for example, the $J=10$ Pade may have better accuracy than the optimized $J=8$. For $J>20$, random error becomes the major issue for the approximations, and reducing the error to less than $10^{-13}$ becomes difficult. Therefore, for double precision usage, $J=20$ is sufficient. The $J=24$ coefficients listed here are for reference, for those who hope to develop more accurate approximations. Due to the random off error, calculating using $p_l$ and $q_l$ would yield better accuracy for large $z$ than using $b_j$ and $c_j$.

It is claimed \cite{Zaghloul2011} that if the term $i\sigma\sqrt{\pi}e^{-s^2}$ is omitted, Eq.(\ref{eq:Ztaylor}) would not match well for the range $y<\sqrt{\pi}x^2e^{-x^2}$ when $x \gg 1$. However, in practical tests, this only slightly affects the imag part of intermediate $x$, and the Weideman method also holds well for $y \simeq 0$.

In Appendix \ref{sec:code}, we provide Matlab and Python code examples of how to use these coefficients to calculate $Z(s)$. In Appendix \ref{sec:weid}, we also provide the coefficients using the Weideman method.

\begin{figure*}
\centering
\includegraphics[width=15cm]{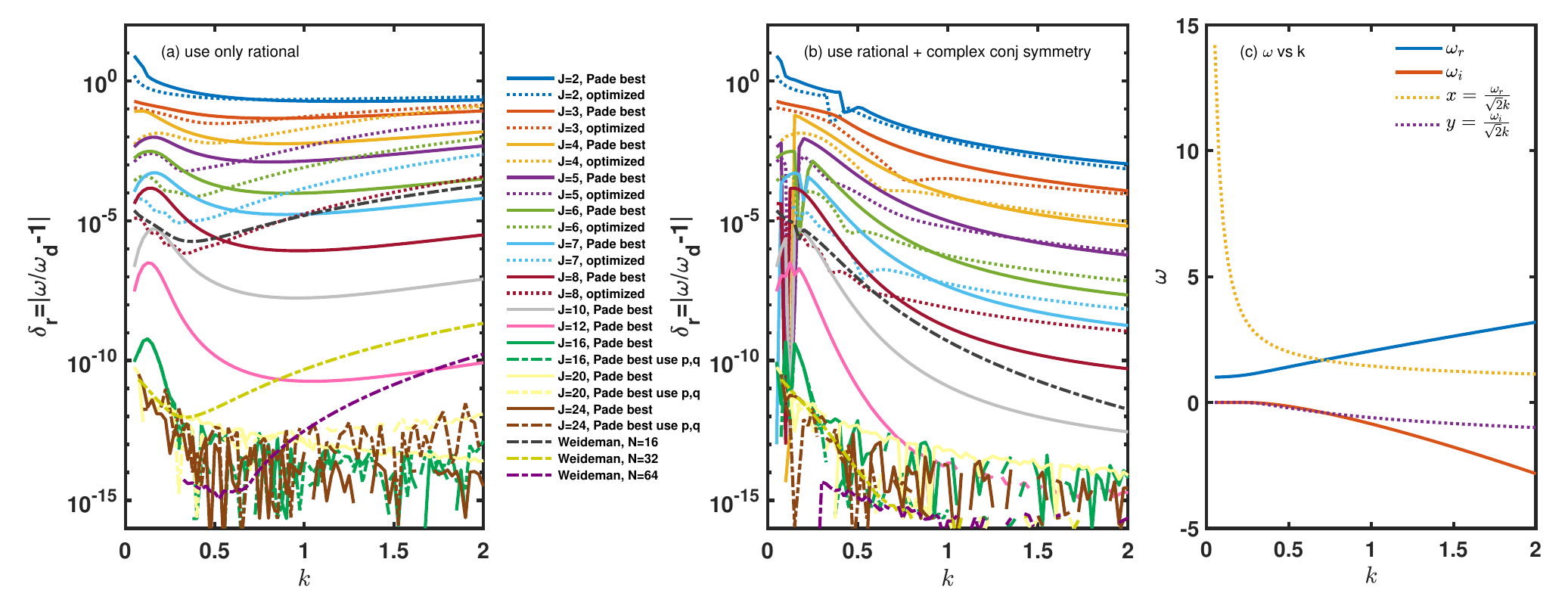}
\caption{Validation of different expansion coefficients through comparison with the Landau damping roots.}\label{fig:landauroots}
\end{figure*}

\begin{figure*}
\centering
\includegraphics[width=15cm]{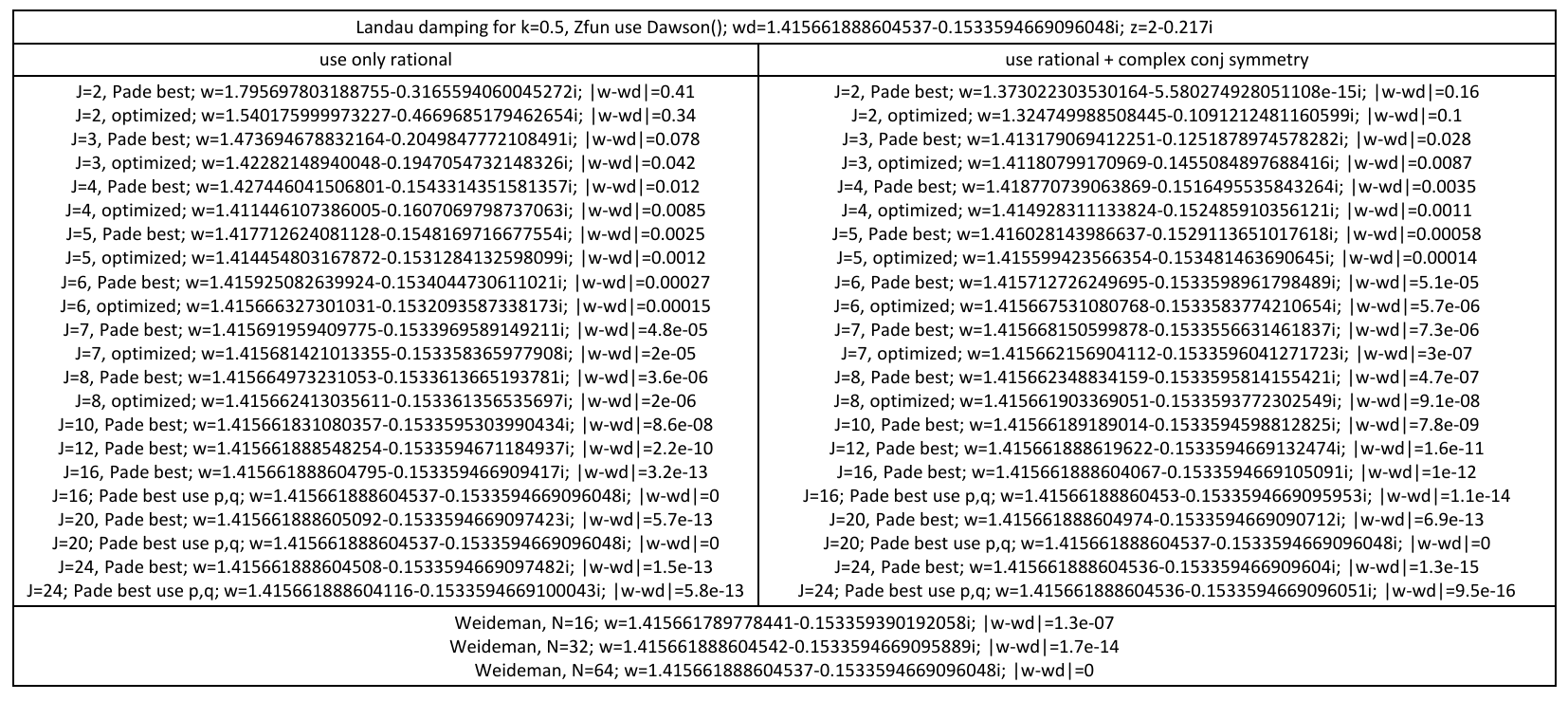}
\caption{Validation of different expansion coefficients through comparison with the Landau damping roots, for $k=0.5$.}\label{fig:landaudata}
\end{figure*}

\subsection{Validation}
The accuracy of $Z$ is not just about the function itself, but rather whether it can accurately capture the physical phenomenon. We employ the one-dimensional electrostatic Landau damping problem to demonstrate the accuracy of the data provided in the table. The problem can be simplified as follows \cite{Xie2013}
\begin{equation}\label{eq:landau}
D(\omega,k)=1+\frac{1}{k^2}[1+zZ(z)]=0,
\end{equation}
with $z=\omega/(\sqrt{2}k)$. For a given $k$, we aim to solve for $\omega$, focusing on the least damping branch.

Results are depicted in Fig. \ref{fig:landauroots} and Fig. \ref{fig:landaudata}. The $J=2$ method cannot be used, as the large error leads to incorrect roots. $J=3,4$ could be viable options for low accuracy calculations. We observed that for the Pade method, larger $J$ generally results in better accuracy across all arguments. However, for the same $J$, optimized coefficients may only control total errors, with local errors potentially increasing, especially for small $s$. Thus, optimization coefficients may not always outperform unoptimized coefficients. This suggests that if only accurate $Z(s)$ is needed (i.e., not aiming to reduce the expansion order), using coefficients from the rational expansion with higher orders (i.e., larger $J$) is preferable over optimizing the coefficients. Hence, we recommend using $J\geq8$ from our table, instead of the older lower $J$ data \cite{Hui1978,Humlicek1979,Humlicek1982}. The coefficients provided here for calculating $Z(s)$ can deliver up to twelve significant decimal digits, limited by double precision data rather than the approach itself. Therefore, the Pade expansion method can compete with Weideman's method and can even be simpler.

Breakdown region is for $k\leq0.15$, where $y<10^{-7}\ll1$, incorrect positive numerical solutions of $\omega_i$ may occur. However, since $\omega_i/\omega_r\ll10^{-8}$ is small, it is less critical to achieve high accuracy for most applications. Humlicek \cite{Humlicek1982} reported that any rational approximation suffers inevitable failure near the real axis. For methods to address this issue, one can refer to \cite{Humlicek1982,Zaghloul2011}.

\section{Summary and Conclusion}\label{sec:summ}
We revisit the issue of rapid calculation of the plasma dispersion function and provide comprehensive rational and multi-pole coefficients for reference. A practical application of this work is to accelerate the computation of the PDRK/BO\cite{Xie2016,Xie2019} code. For instance, an optimized $J=6$ can achieve the same maximum error as the former Pade $J=8$ method, with a complexity of $O(J^{2.7})$, resulting in a speedup of approximately 2.1 times. The optimized $J=8$ method also reduces the maximum error by around two orders of magnitude (80 times) compared to the usually used Ronnmark $J=8, I=10$ Pade coefficients. However, if reducing the order is not necessary, we recommend using larger $J$ values (such as $J=10,12,16,20,24$) to improve global accuracy.
The major possible inaccuracy occurs at small $y$ for intermediate/large $x$, which is found to be not a significant issue for most practical applications.

The demonstration of errors of $Z(s)$ itself and its application to the Landau damping problem also provide a validation range for different coefficients and offer guidance for further selection.


\begin{figure*}
\centering
\includegraphics[width=15cm]{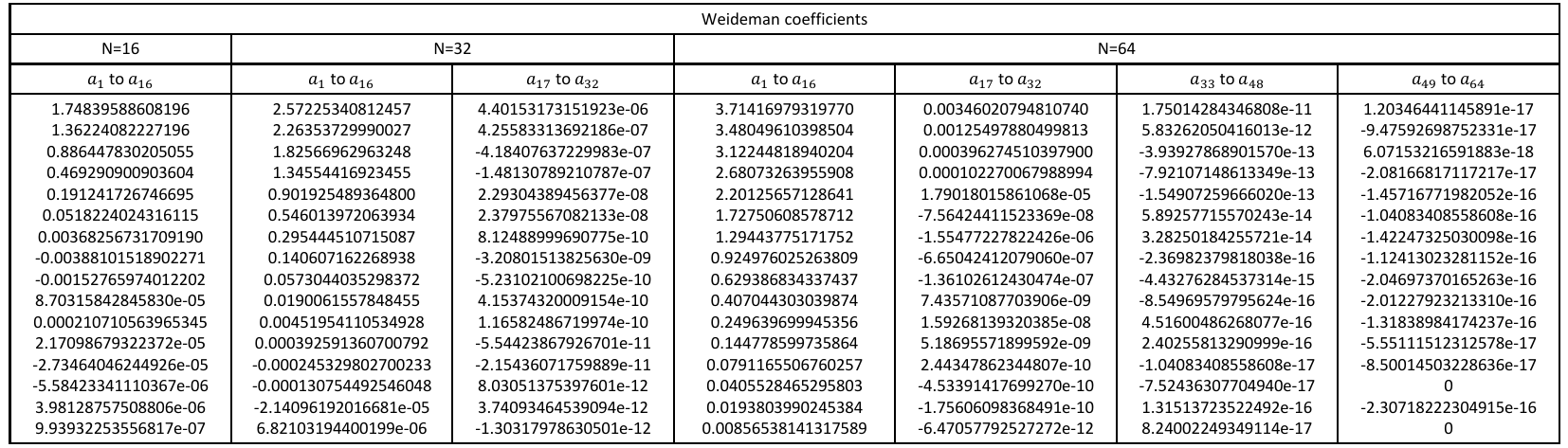}
\caption{Weideman coefficients for $N=16$, $N=32$, and $N=64$ for the calculation of the $Z$ function.}\label{fig:weidemandata}
\end{figure*}

\appendix
\section{Weideman coefficients}\label{sec:weid}
The Weideman method for calculating $Z$ relies on the Fast Fourier Transform (FFT), which may be less convenient for coding, such as in Fortran. However, the FFT coefficients can be separated out for practical applications. We provide coefficients for $N=16$, $N=32$, and $N=64$ here, which could be used for high accuracy computation of $Z$ for most ranges. The expansion is given by \cite{Weideman1994}
\begin{equation}\label{eq:WeidZ}
Z(z)\simeq\frac{i}{L-iz}+\frac{2i\sqrt{\pi}}{(L-iz)^2}\sum_{n=0}^{N-1}a_{n+1}\Big(\frac{L+iz}{L-iz}\Big)^n,~y\geq0,
\end{equation}
with $L=2^{-1/4}N^{1/2}$. The above form also holds for weak damping case, i.e., $y<0$, but not too far from the real axis as demonstrate in Figs.\ref{fig:zJpoleerr_y=-0.1}, \ref{fig:zJpoleerr_y=0}, and \ref{fig:zJpoleerr_y=0.1}. For accurate calculation when $y < 0$, we can still use Eq. (\ref{eq:sys}). The results are shown in Fig. \ref{fig:weidemandata}.

Note that this method can also be used for non-Maxwellian distributions directly \cite{Xie2013}.

\begin{figure*}
\centering
\includegraphics[width=15cm]{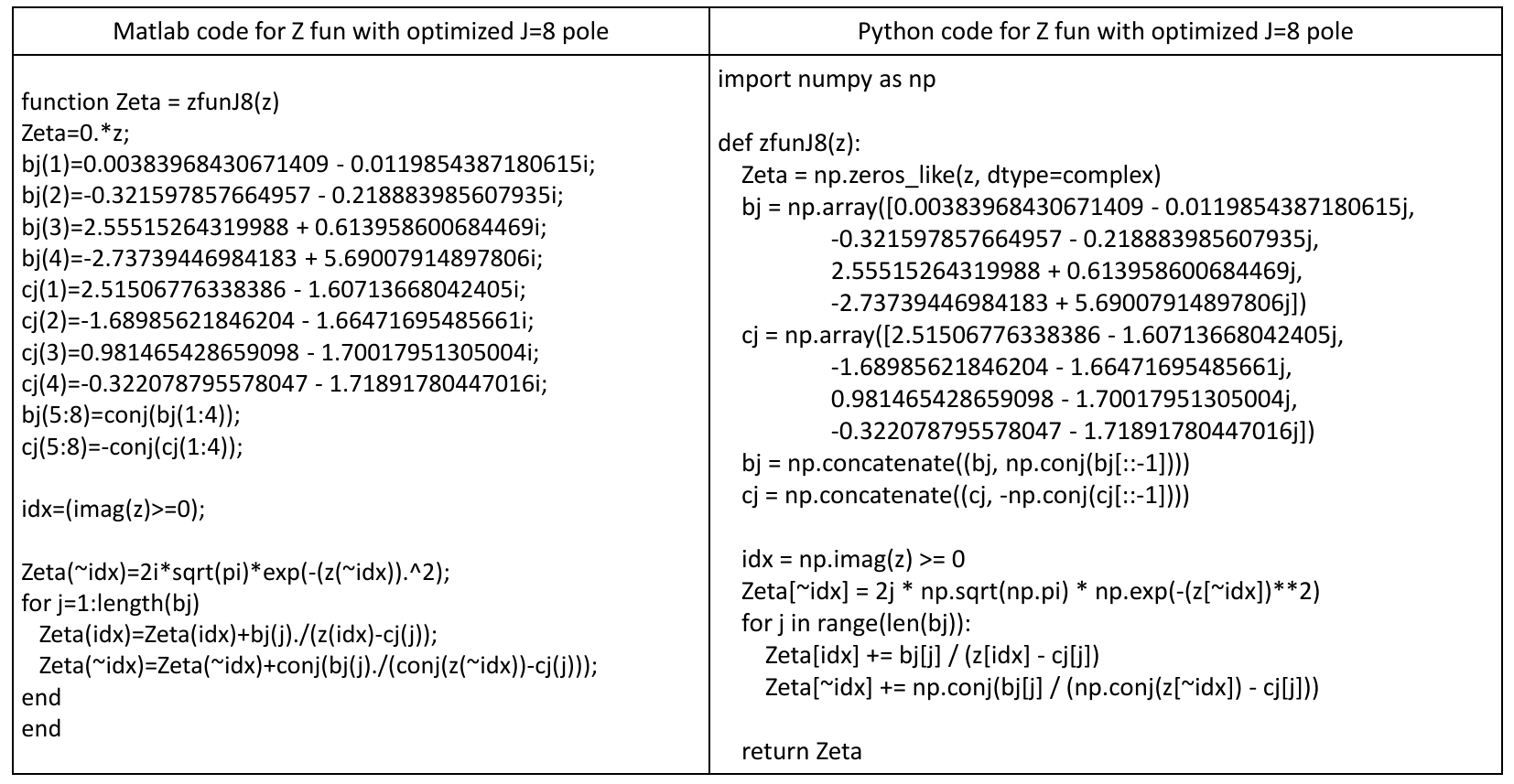}
\caption{Sample code for calculate $Z$ function with optimized $J=8$ pole for all range of argument $z$, with max errors of $10^{-6}$. One who needs higher accurary, can use the larger $J$ coefficients, such as $J=10,12,16,20,24$.}\label{fig:samplecode}
\end{figure*}

\section{Short code example}\label{sec:code}
It is usually surprising to the beginners that the simple several rational terms can calculate the complicated complex integral function $Z(s)$ to high accurate. We thus also provide sample codes in Fig.\ref{fig:samplecode} for easy access, which is valid for $z$ at entire plane.

\end{document}